\documentclass[referee]{aa-mpa}
\usepackage{graphicx}
\makeatletter 
\let\@internalcite\cite
\def\cite{\@ifstar{\citeyear}{\citefull}}
\def\citefull{\def\astroncite##1##2{##1 ##2}\@internalcite}
\def\citeyear{\def\astroncite##1##2{##2}\@internalcite}
\def\citeau{\def\astroncite##1##2{##1}\@internalcite}
\def\citen{\def\astroncite##1##2{##1 (##2)}\@internalcite}
\def\possesivcite{\def\astroncite##1##2{##1's (##2)}\@internalcite}
\def\@citex[#1]#2{\if@filesw\immediate\write\@auxout{\string\citation{#2}}\fi
  \def\@citea{}\@cite{\@for\@citeb:=#2\do
    {\@citea\def\@citea{; }\@ifundefined
       {b@\@citeb}{{\bf ?}\@warning
       {Citation `\@citeb' on page \thepage \space undefined}}%
{\csname b@\@citeb\endcsname}}}{#1}}
\def\@cite#1#2{#1\if@tempswa , #2\fi}
\def\@biblabel#1{}
\makeatother

\begin{document}

\thesaurus{09(02.14.1, 02.16.1, 06.07.1, 06.09.1)}

\title{Solar models and electron screening}

\author{A.~Weiss\inst{1,2,3} \and M.~Flaskamp\inst{1} 
\and  V.N.~Tsytovich\inst{4}}

\institute{Max-Planck-Institut f\"ur Astrophysik,
           Karl-Schwarzschild-Str.~1, 85748 Garching, \\
           Federal Republic of Germany
           \and
           Princeton University Observatory,
           Peyton Hall, Princeton NJ~08540, USA
           \and 
           Institute for Advanced Study, Einstein Dr.,
           Princeton NJ~08540, USA
           \and 
           Dep.\ of Theoretical Physics, General Physics Institute,
           Russian Academy of Sciences, Moscow, Russia
           }

\offprints{A.~Weiss; (e-mail: weiss@mpa-garching.mpg.de)}
%\mail{A.~Weiss}

\date{Received; accepted}

%\authorrunning{Weiss \&}
%\titlerunning{A new paper}

\maketitle

\begin{abstract}
We investigate the sensitivity of the solar model to changes in the
nuclear reaction screening factors. We show that the sound speed
profile as determined by helioseismology certainly rules out changes
in the screening factors exceeding more than 10\%. A slightly
improved solar model could be obtained by enhancing screening by about
5\% over the Salpeter value. We also discuss
how envelope properties of the Sun depend on screening, too. We
conclude that the solar model can be used to help settling the
on-going dispute about the ``correct'' screening factors.
\end{abstract}

\vspace{4.0cm}
\centerline{\sl accepted for publication in Astronomy \& Astrophysics}

\clearpage

\section{Introduction}

The solar interior is, thanks to the tremendous progress of
helioseismology, known to such great accuracy that the Sun can in fact
be used as a laboratory for physics. Examples are investigations about
the equation of state (\cite{dana:00}) or axion properties
(\cite{swr:99}). In this paper, we will apply it to the 
Coulomb screening of the nuclear reactions. 

The plasma correction to nuclear reaction rates, also known as
screening, is one of the ingredients for stellar and solar model
calculations, which has repeatedly been rediscussed in the
literature. The standard derivation by \citen{salp:54} discusses {\em
electrostatic screening} in the Debye-approximation, where the
electrostatic potential around an ion of charge $Z_1e$ is
\begin{equation}	
\Phi = (Z_1e/r)\exp (-r/D),
\label{e:1}
\end{equation}
where $D$ is the Debye radius ($D=kT / 4\pi e^2 n_e$; $n_e$:
electron density), within which ($r\ll D$) the electrostatic
potential around $Z_1$ is reduced to 
\begin{equation}	
\Phi = Z_1e/r - Z_1e/D.
\label{e:2}
\end{equation}
This reduced Coulomb potential gives rise to an increased reaction
rate by a factor
\begin{equation}
f = \exp \left( Z_1Z_2e^2\over kTD \right) \approx 1+f_\mathrm{S} =
1+0.188Z_1Z_2 \xi^{1\over 2} \rho^{1\over 2}T_6^{-{3\over 2}}, 
\label{e:3}
\end{equation}
$Z_2e$ denoting the partner charge in the nuclear reaction, 
and $\xi=\sum_Z {(Z^2+Z)X_Z\over A_Z}$ the effective charge.

Eq.~\ref{e:3} has been challenged in various papers
(e.g.\ \cite{csk:88}; \citeau{shsh:96} \citeyear{shsh:96},
\citeyear{shsh:00}; \cite{sav:99}), either developing
different pictures for the configuration of $Z_2$ around $Z_1$, or by
pointing out the static character of Salpeter's derivation, asking for
a dynamical one, since we are dealing with a plasma.
All these papers have been refuted by other work pointing out flaws,
inconsistencies or by re-derivation of Eq.~\ref{e:3} (e.g.\
\cite{bg:97}, \cite{gb:98}). Recently, \citen{bbgs:00} have summarized
five different derivations of the plasma correction to nuclear
reaction rates at the centre of the Sun, all arriving at the Salpeter
formula. At the same time they review several papers with deviant
screening rates, stressing that all these papers arrive at different
factors. In their summary, \citen{bbgs:00} emphasize that in the
future proving the correctness of screening formulae
should be part of the corresponding work. Among the work criticised
was also that of one of the present authors (\cite{tb:00};
\cite{tsyto:00}), in which screening results in a {\em suppression} of
nuclear reaction rates in the Sun.

Since many of the paper in this field are not easy to understand and
in particular their correctness not easy to be verified nor falsified,
we used in \citen{fwt:00} a different approach. Following our
previous applications of the Garching Solar Model (GARSOM;
\cite{schl:99}) as a highly sensitive laboratory for stellar and
particle physics (e.g.\ \cite{schl:99}; \cite{swr:99}) we applied the
screening rates of \citen{tsyto:00} to our solar model and compared the
resulting sound speed with a model with Salpeter's equation. In this
preliminary work we reported about
changes in sound speed both at the solar centre and in the radiative
regions around 
$r/R_\odot \approx 0.65$ ($R_\odot$ is the solar radius), which are
inconsistent with helioseismological results.

In the present paper we are extending our analysis to general
modifications of Eq.~\ref{e:3} and repeat it for a slightly modified
derivation of the screening compared to the original work by
\citen{tb:00}. In Section~2 
we will briefly summarize our solar models and describe the modified
screening values. In Section~3 we show the influence on sound speed,
neutrino rates and global solar properties and demonstrate that, given
the accuracy of the helioseismologically inferred sound speed, even the
case of reactions rates without screening ($f=1$) can be
excluded.\footnote{We note that a very similar investigation has
recently and independently been published by \citen{fbv:00}.}
We
then turn around our approach to show that a slightly improved solar model can be
obtained by increasing Salpeter's screening by a moderate amount
($\approx 5\%$). Finally, in the last section, we conclude that the
present work demonstrates 
convincingly that the Sun can be used as a laboratory for nuclear
reaction screening; this could be helpful in the future to settle the
on-going dispute about screening.

\section{Calculations and models} 

\subsection{Solar models}

We calculated solar models with the program and input physics as
described in \citen{schl:99} or \citen{swr:99}. The physics is
very similar to other standard models such as published
recently by \citen{bpb:00}. We avoid repeating details here, but
emphasize that helium and metal diffusion are included and the present
solar luminosity, radius and surface value of $Z/X = 0.0245$ are
matched (note that \cite{bpb:00} use a slightly lower value of
0.0230; \cite{gs:98}). The resulting model parameters (Table~\ref{t:1})
and sound profile (Fig.~\ref{f:1}; solid line) agree well with
\citen{bpb:00}.  

For our standard model (``GARSOM'') we have used Salpeter's
formula. In all other models, screening factors were
replaced as described next. Otherwise the model calculations are
identical.

\subsection{Screening factors}

Except for the {\em Tsytovich-screening} model (Fig.~\ref{f:1};
dashed line), the screening 
factor $f$ of all reactions of pp-chains and CNO-cycle was
multiplied by a constant factor $g$, ranging 
from 0.9 to 1.1, such that $f=(1+f_\mathrm{S})\cdot g$
for the models shown in Fig.~\ref{f:2}, or was replaced by $f=1$ for
the {\em no-screening} model (Fig.~\ref{f:1}; dash-dotted line).

% we also have the case "no screening" nur fuer pp
% and Salpeter in CNO - directly compared to Tsytovich case
% in this case, T get's higher, CNO is favored (T and screening)
% and fluxes go up!

For the {\em Tsytovich-screening} model a varying screening factor
depending on reaction and composition was used for the $\mathrm{p}
+\mathrm{p}$, $^3\mathrm{He}+^3\mathrm{He}$, $^3\mathrm{He}+
^4\mathrm{He}$, $^7\mathrm{Be}+\mathrm{p}$, and $^7\mathrm{Be}+e^-$
reactions.
(For all other
reactions standard Salpeter values were used.)  These factors were 
determined by starting out with the formulae given in \citen{tb:00},
which involve the solution of multi-dimensional integrals, which
depend on temperature, density, and abundances of the participating
nuclei. Instead of evaluating these integrals at all points within the
solar model, we did so for a selected number of mass coordinates for
several models along a standard model evolution sequence and found that to
very good approximation the screening factors can be expressed as a
function of helium content (as a parameter describing both the spatial
coordinate and the evolution with time within the energy generating
core). In Eq.~\ref{e:3} we replace $Z_1Z_2\xi^{1\over2}$ by $F\cdot
L_{ij}$, where $F$ is depending on composition and $L_{ij}$ on
reaction ${ij}$. For the centre of the Sun we recover the screening
factors given in Table~1 of \citen{tsyto:00} when using $F$ and
$L_{ij}$ from \citen{tb:00}.

We note that the original results of \citen{tb:00} have been modified
slightly by V.N.~Tsytovich until the time of the computations 
presented here and are subject to further ongoing research. With
respect to the original paper, the quantity $L_{ij}$ became
temperature-dependent (except for $e^-$-capture on $^7$Be) and some
additional terms in expansion series had been added. Meanwhile the
formalism has been improved to treat two-particle distribution
functions, a sign-error in one term has been corrected for, and it
appears that in addition to the dynamical screening terms the static
ones, which correspond to the Salpeter-screening and which canceled in
the original derivation (\cite{tb:00}) could remain, depending on
whether plasma perturbations induced by the nuclear reactions
themselves can decay in time.

The new derivation and results will be published in a
forthcoming paper (Tsytovich 2001, in preparation). 
We therefore emphasize that the purpose
of the present paper is not to investigate the \citen{tsyto:00}
screening rates, because they are already under further development,
but to demonstrate the capability of the Sun as a laboratory for
screening using the screening formulation by \citen{tsyto:00} as an
example.

\section{Results}

\subsection{Sound speed}

\begin{center}	
\begin{figure}
\includegraphics[draft=false,scale=0.55,angle=90]{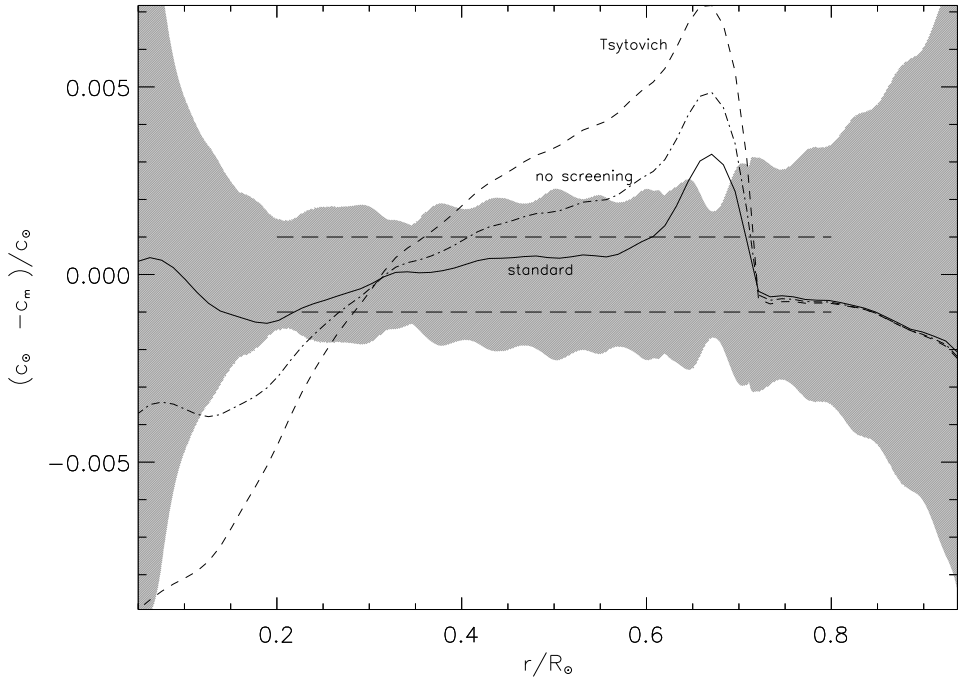}
\caption[]{Comparison of the sound speed profile for our three models
using Salpeter- (solid line), Tsytovich- (dashed) and no screening
(dash-dotted) with the sound speed derived from helioseismology (see
text). The shaded area and the long-dashed line refer to different
error estimates in the latter} 
\protect\label{f:1}
\end{figure}
\end{center}

\begin{center}
\begin{figure}
\includegraphics[draft=false,scale=0.55,angle=90]{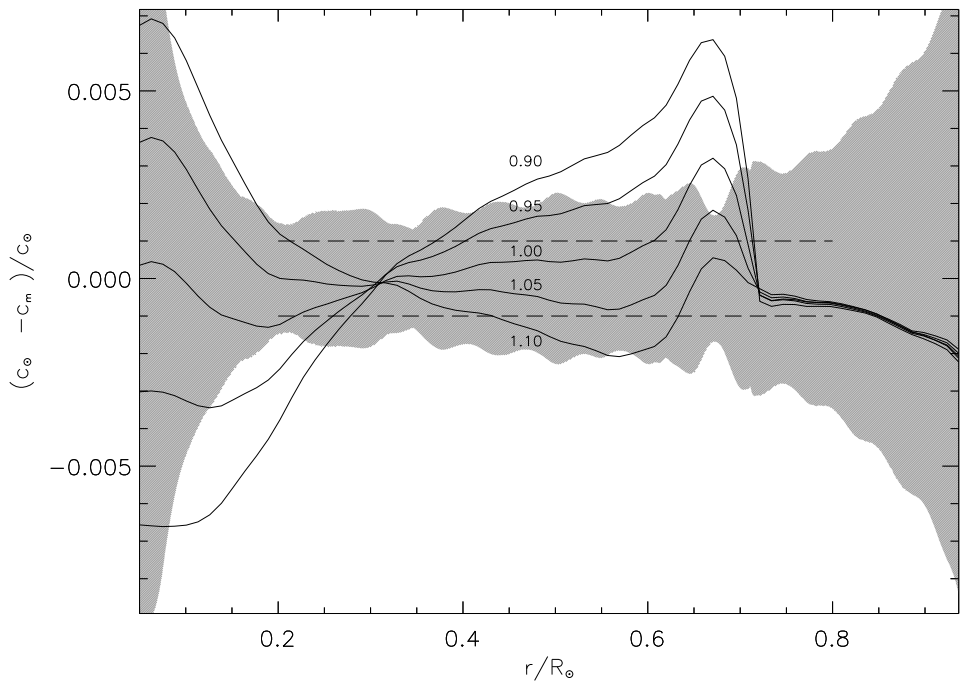}
\caption[]{As Fig.~\ref{f:1}, but for models with the Salpeter
screening factor $f=1+f_\mathrm{S}$ (Eq.~\ref{e:3}) multiplied by a
constant factor indicated along the various lines}
\protect\label{f:2}
\end{figure}
\end{center}

The results of our experiments are displayed in Figs.~\ref{f:1} and
\ref{f:2} for the sound speed difference between models
($c_\mathrm{m}$) and the seismic Sun ($c_\mathrm{s}$). The inferred
sound speed profile we took from \citen{basu:98}. For the error range
of $c_\mathrm{s}$ we show two different results: 
the extremely conservative estimation by \citen{ddf:97}, shown
as the grey-shaded area in Fig.~\ref{f:1}, and a more recent
analysis about the uncertainty of the inversion procedure
(\cite{bpj:00}). In the latter paper use of SOHO-results was made and
the authors concluded that the relative
sound speed errors due to the measurements are of order $3\cdot 10^{-4}$
($3\sigma$)  and that inversion method and starting
model add about equally, thus that a conservative error range based on
this paper would be of order $1\cdot 10^{-3}$ for $0.2 < r/R_\odot <
0.8$. This is indicated by the long-dashed line in both figures.

It is immediately clear from Fig.~\ref{f:1} that the model with {\em
Tsytovich-screening} is stronly discrepant with the seismic model and
even the {\em no-screening} model is outside the conservative error
range for small radii and below the convective envelope. The deviation
from the standard model is of order the conservative error range or
larger, such that it can be excluded. This is in particular true for
the more up-to-date error estimate by \citen{bpj:00} in the radiative
interior. 

It is interesting to note that going from {\em Salpeter-} to {\em
Tsytovich-screening} there are systematic changes in both the core
and sub-convective regions, which are anti-correlated. The model's
sound speed is increasing in the core due to the higher central
temperatures 
necessary to provide the same number of pp-reactions per second (the
{\em solar luminosity constraint}).
At the same time $c_\mathrm{m}$ is decreasing in
that part of the radiative zone outside the energy-generating core
(extending to $r/R_\odot \approx 0.2$) due to a slightly more expanded
structure. The Sun appears to be more centrally concentrated than it
is for standard screening. Recognizing this it is
only a small step to ask if the two extrema in Fig.~\ref{f:1} along
the {\em standard}-model can be reduced simultaneously by {\em
increasing} the screening above the Salpeter value.

The result of the corresponding experiment is shown in Fig.~\ref{f:2}
for different factors 
multiplied to Eq.~\ref{e:3} for all reactions. An increase of the
screening can indeed improve the situation at the discrepant bump
around $r/R_\odot \approx 0.65$, but leads to a comparatively large
change in the core. An increase over Salpeter's screening by more than
10\% will become problematic even in the case of the conservative
error estimates; 
reductions by 5\% or more can clearly be excluded. There are two
lessons to be drawn from this:

\noindent 1. agreement with the seismic sound speed profile
is obtained from screening factors not deviating by more than +10/-5\%
from Salpeter's formula;

\noindent 2. the discrepant region below the convective envelope
might not necessarily be due to missing physics {\em only} in that
region; it should not be forgotten that the Sun is a system coupling
different regions and that changes in the core also affect the rest of
the Sun.

We close this section by reminding that the structural changes just
discussed depend to first order only on the screening of the
pp-reaction (as we have verified in test calculations), which is the
one directly connected to the solar luminosity. All statements made
above therefore restrict the screening of only this reaction. We also
remind the reader that the reaction rate itself is influencing the
model as shown by \citen{ac:99}; we are using the rate by
\citen{adel:98}, where $S(0) = 4.00\cdot 10^{-25}$~MeVb. This value,
together with that of the central metallicity ($Z_c=0.021$) is
marginally consistent with helioseismic data, as shown in
\citen{ac:99}, Figure~2.

\subsection{Other properties}

\begin{table}
\caption[]{Properties of the three solar models of
Fig.~\ref{f:1}. In the upper part neutrino fluxes (in
$\mathrm{s}^{-1} \cdot \mathrm{cm}^{-2}$) as predicted from the
models are given. In the central 
part the corresponding expected measurements in the three experiments 
are listed; the measured values are given in the first column. For the
Cl- and Ga-experiments the unit is in SNU, for Super-Kamiokande it is
in units of the standard model prediction. The lower part shows some
global properties of the initial 
and present Sun. Note that our mixing-length parameter is different
from those of other standard models because of the different treatment
of convection and atmospheres in our models (see \cite{schl:99}).}
\protect\label{t:1}
\begin{center}
\begin{tabular}{l|r|r|r}
	& Salpeter & no screening  & Tsytovich  \\
\hline%\noalign{\smallskip}       
 pp 	& $5.93\cdot 10^{10}$&$5.96\cdot 10^{10}$&$5.96\cdot 10^{10}$\\
 pep  	& $1.390\cdot 10^{8}$&$1.427\cdot 10^{8}$&$1.463\cdot 10^{8}$\\
 hep 	& $2.076\cdot 10^{3}$&$2.107\cdot 10^{3}$&$2.456\cdot 10^{3}$\\
 7Be 	& $4.812\cdot 10^{9}$&$4.677\cdot 10^{9}$&$4.023\cdot 10^{9}$\\
 8B  	& $5.054\cdot 10^{6}$&$5.353\cdot 10^{6}$&$5.225\cdot 10^{6}$\\
 13N  	& $5.849\cdot 10^{8}$&$5.004\cdot 10^{8}$&$8.101\cdot 10^{8}$\\
 15O 	& $5.072\cdot 10^{8}$&$4.206\cdot 10^{8}$&$7.323\cdot 10^{8}$\\
 17F	& $6.251\cdot 10^{6}$&$5.017\cdot 10^{6}$&$9.200\cdot 10^{6}$\\
% 13N if CNO is Salpeter, pp is no-screening:   6.705e8
% 15O                                           6.920e8
% 17F                                           7.374e6
\hline%\noalign{\smallskip}
Sum &$6.54\cdot 10^{10}$ & 
   $6.54\cdot 10^{10}$&   $6.54\cdot 10^{10}$ \\
\hline%\noalign{\smallskip}
%Cl (2.56)& $7.579$ & $6.748$ & $3.717$ \\
%Ga (72.5)& $128.40$ & $124.79$ & $119.03$ \\
%SK (47.8\%) & 100\%& 87.1\%& 32.6 \\
Cl (2.56)& $7.579$ & $7.819$ & $7.790$ \\
Ga (72.5)& $128.40$ & $127.05$ & $127.61$ \\
SK (47.8\%) & 100\%& 105.9\% & 103.4\% \\
% the three values for the CNO-Salpeter/pp-no case
% Cl 8.00, Ga 129.72, SK 1.07
\hline%\noalign{\smallskip}
$\alpha$	&0.9750  &0.9281  &0.8731 \\
$Y_\mathrm{i}$	&0.2747  &0.2734  &0.2721 \\
$Z_\mathrm{i}$	&0.0199  &0.0200  &0.0201 \\
$Y_\mathrm{s}$	&0.2448  &0.2428  &0.2403 \\
$Z_\mathrm{s}$	&0.0181  &0.0181  &0.0182 \\
$R_\mathrm{bcz}/R_\odot$ &0.7135  &0.7163  &0.7188 \\ 
$T_\mathrm{c}$	&$1.5707\cdot 10^{7}$&$1.5792\cdot 10^{7}$&
	$1.5947\cdot 10^{7}$   \\
\end{tabular}
\end{center}
\end{table}

The Sun does not only show the influence of screening in its sound
speed profile, but also in some other properties (Table~\ref{t:1};
lower part). Related to results 
based on helioseismology are the depth of the convective envelope and
the present surface helium content, for which the determined values
are $R_\mathrm{bcz}/R_\odot = 0.713\pm0.001$ (\cite{ba:97}) and
$Y_\mathrm{s} = 0.249\pm 0.003$ (\cite{ba:95}). While the
{\em Salpeter}-model nicely fits, the {\em no-screening}
case already is at the border of the error range and the {\em
Tsytovich}-model disagrees with the value for $R_\mathrm{bcz}$ and is
only marginally consistent with that for $Y_\mathrm{s}$. Note that
these are envelope resp.\ surface quantities, but the change to the
physics is done deep inside the core! 

The central temperature $T_\mathrm{c}$ is increasing with decreasing
reaction efficiency, as expected. This is due mainly to the
pp-reaction, which is the one closely linked to the luminosity of
the Sun and therefore reduced screening must be balanced by higher
temperature. The effect is measurable because of the rather low
dependence of the reaction on temperature under solar conditions. For the same
luminosity reason the flux of pp-neutrinos remains almost unchanged
(Table~\ref{t:1}; upper part). Temperature sensitive $\nu$-emission
rates should increase more ($pep$ and $hep$), unless the influence of
the reduced screening is dominant. Considering the $^7$Be flux, the
transition to {\em no-screening} reduces it by 2.7\%, but the
further step to {\em Tsytovich-screening} leads to a reduction of
16.3\%. The $^8$B-source, however, actually increases due to the
increased temperature and its extreme temperature sensitivity. Note
that in \citen{fwt:00} we reported about a strongly reduced
$^8$B-flux. This result was wrong, because we erroneously had not
modified the screening of the $^7\mathrm{Be}+e^-$ reaction, which does
not appear explicitely in our nuclear network, but only as a rate
relative to the $^7\mathrm{Be}+p$-capture. 

%The CNO-$\nu$-fluxes require special attention. The screening for them
%has been omitted in the {\em no-screening} case, therefore the rates
%tend to be reduced. There are no screening factors available from
%\citen{tsyto:00} for them, so Salpeter's formula was used, and the
%effect of increased temperature is dominant. 

As a consequence of the modified  neutrino emission rates, the
predictions for the neutrino experiments change. Most interesting is
the case of Super-Kamiokande, which the {\em Tsytovich-screening} model
predicts to measure even more neutrinos than the standard model does,
although the rate is much stronger suppressed (see \cite{tsyto:00},
Tabel~1) than the pp-reaction. However, due to the luminosity
constraint and increased central temperature, the modified screening
effect is overcompensated.

\section{Conclusions}

We have demonstrated that the Sun has become a sensitive laboratory to
investigate the correct plasma screening for its own nuclear
reactions. The high precision of the seismic Sun does not allow for
deviations of more than a few percent from the original formula by
\citen{salp:54}. A slightly enhanced screening factor improves the
agreement with the seismic sound speed profile.
We emphasized that not only the central parts, where
the influence of changes to the nuclear reaction rates is immediate,
but also the outer layers will have a modified structure pointing out
discrepancies. We have to recall, however, that this is mostly due to
changing the screening of the pp-reaction. Reactions in the higher
pp-chains only will not influence the 
models significantly. Because of these reaction rates, in contrast, the
neutrino fluxes will
change, not leading to the once proposed solution of the
classical neutrino problem (which, as is known by now, cannot be
solved by changing reaction rates only; \cite{hbl:94}), but instead
to even larger discrepancies, as we found for the Super-Kamiokande
prediction of the {\em Tsytovich-screening} model, although this
modified screening suppresses the reactions at fixed temperature.

One should keep in mind that physical pictures deriving alternative
screening rates should in principle also be used to rederive the
configurational effects in the equation of state and opacities to
arrive at consistent physical model input.  Our present work thus must
remain incomplete in this respect, but we trust that the success
of the standard solar model with Salpeter's screening is unlikely to
be due to a conspiracy of using erroneous physics for three major
model ingredients.

We believe that the Sun itself will help to settle the dispute about
the true amount of screening of nuclear reactions. It has become a
truely invaluable celestial laboratory for physics.

\begin{acknowledgements}
We thank H.~Schlattl for his assistance in calculating the solar
models and for permission to use his excellent code.
A.W.\ acknowledges travel support from the Fulbright foundation and is
grateful for the hospitality at
the Princeton Observatory and the Institute for Advanced Study. He
thanks A.~Gruzinov, P.~Krastev, and J.~Bahcall for stimulating
discussions. This work was supported in part by 
``Sonderforschungsbereich 375-95 f\"ur Astro-Teilchenphysik
der Deutschen Forschungsgemeinschaft''.
\end{acknowledgements}
\bibliographystyle{aa_weiss}
\bibliography{screening,sun,weiss}

\end{document}